\documentclass[%
 reprint,
 superscriptaddress,
 amsmath,amssymb,
]{revtex4-2}

\usepackage{bbold}
\usepackage{mathptmx}
\usepackage{subfig}
\usepackage{psfrag,graphicx}
\usepackage{dcolumn}
\usepackage{bm}
\usepackage{latexsym}
\usepackage{epstopdf}
\usepackage{color}
\usepackage[english]{babel}
\usepackage{amsfonts}
\usepackage{appendix}
\usepackage[justification=justified]{caption}
\usepackage{aeguill}

\definecolor{mygrey}{gray}{0.35}
\definecolor{myblue}{rgb}{0.2,0.2,0.8}
\definecolor{myzard}{cmyk}{0,0,0.05,0}
\definecolor{mywhite}{rgb}{1,1,1}
\definecolor{mywhite}{rgb}{1,1,1}
\definecolor{myred}{rgb}{1,0.,0.3}

\usepackage[colorlinks=true,citecolor=myblue,linkcolor=myblue]{hyperref}

\def\ba{\begin{align}}
\def\enda{\end{align}}
\def\bi{\begin{itemize}}
\def\ei{\end{itemize}}

\def\be{\begin{equation}}
\def\ee{\end{equation}}
\def\bea{\begin{eqnarray}}
\def\eea{\end{eqnarray}}
\def\bse{\begin{subequations}}
\def\ese{\end{subequations}}

\newcommand{\ket}[1]{|{#1}\rangle}                       

\newcommand{\Ignore}[1]{ }

\def\i{\text{i}}

\begin{document}

\preprint{APS/123-QED}

\title{Spin-chain-star systems: entangling multiple chains of spin qubits}

\author{R. Grimaudo}
\address{Dipartimento di Fisica e Chimica ``Emilio Segr\`{e}",
Universit\`{a} degli Studi di Palermo, viale delle Scienze, Ed. 18, I-90128, Palermo, Italy}


\author{A. S. M. de Castro}
\address{Universidade Estadual de Ponta Grossa, Departamento de F\'{\i}sica, CEP 84030-900, Ponta Grossa, PR, Brazil}

\author{A. Messina}
\address{ Dipartimento di Matematica ed Informatica, Universit\`a degli Studi di Palermo, Via Archirafi 34, I-90123 Palermo, Italy}

\author{D. Valenti}
\address{Dipartimento di Fisica e Chimica ``Emilio Segr\`{e}",
Universit\`{a} degli Studi di Palermo, viale delle Scienze, Ed. 18, I-90128, Palermo, Italy}


\date{\today}

\begin{abstract}
We consider spin-chain-star systems characterized by $N$-wise many-body interactions between the spins in each chain and the central one.
We show that such systems can be exactly mapped into standard spin-star systems through unitary transformations.
Such an approach allows the solution of the dynamic problem of an $XX$ spin-chain-star model and transparently shows the emergence of quantum correlations in the system, based on the idea of  entanglement between chains. 
\end{abstract}



\maketitle

\section{Introduction}

Superposition and entanglement are at the basis of the remarkable
advantages of using quantum mechanics over classical
physics in quantum information science \cite{Yung19,Terhal18,Harrow17}.
Such quantum resources are exploited in several technological applications ranging from quantum simulation \cite{Kokail21,Dalmonte18}, quantum metrology \cite{Carollo18, Carollo19, Dobrza14, Riedel10, Joo11} and quantum cryptography \cite{Schimpf21,Yin20} to quantum computing algorithms \cite{Bruss11,Zidan18}.
Besides entanglement, other quantities have been discovered and proposed as quantum resources over the last years, such as, for example, discord \cite{Ollivier01,Henderson01}, coherence \cite{Baumgratz14}, steering \cite{Cavalcanti14}, and contextuality \cite{Grudka14}.

Generating entanglement and superposition states of large systems is a target which has kindled a growing interest in physical scenarios characterized by small quantum systems through which it is possible to control and coherently manipulate mesoscopic
environments \cite{Dong19,Villazon19,Koch16}.
In this sense, in recent decades, a great attention has been payed in studying spin systems which have been successfully applied in quantum information \cite{Bennett00,Das13,Troiani11}.
The simplest imaginable setting consists in a single spin-qubit (either just a qubit, or a spin-1/2 or, more generally, a two-level system) that controls other $N$ spin-1/2's homogeneously distributed in a circle centred on it.
This system is commonly known as central spin system \cite{Hutton04} and the Hamiltonian models used to describe it are called spin-star models.
In such a star-shaped system the $N$ `environmental' spins do not interact with each other directly, but only with the central one which, hence, plays the role of a bridge through which quantum correlations between the spins surrounding it can arise.

The interest towards central spin models has remarkably grown thanks to: I) their suitability in describing the hyperfine interaction in quantum dots \cite{Urbaszek13} and the interactions between nuclear spins and nitrogen-vacancy centers in diamond \cite{Schwartz18,London13}; II) their broad applicability in different fields like quantum information \cite{Yung11}, quantum metrology and sensing \cite{Sushkov14,He19}, quantum thermodynamics \cite{Arisoy21} and fundamental aspects \cite{Haddadi21}.
Further, a lots of works have been developed to investigate the quantum correlation and thermal entanglement arising among spins in the star framework \cite{Byrnes19,Haddadi19,Xu18,Anza10,Militello11}, 
as well as the Markovian and non-Markovian dynamics induced by a surrounding bath interacting with the spin-star system \cite{Motamedifar19,Wang13,Ferr08}.

Recently, the idea of spin-chain-star system has been proposed \cite{Ping13, Yao11} where the control spin occupies the center of a star of $M$ rays (chains), arranged at angular distance $2 \pi / M$, each of which hosts the same sequence of $N$ spins, generally equidistant.
There is no interaction between different chains and the spins in a given radius may not even be the same \cite{Zhu18}.
The spin-spin interactions within each chain and with the control spin are described by Hamiltonian terms strictly related to the physical scenario to be studied.
In \cite{Eoghan21}, for example, the same system has been analysed to study the effects related to the quantum darwinism in such a structured environment.

\begin{figure*}[htp] 
\begin{center}
\subfloat[][]{\includegraphics[width=0.35\textwidth]{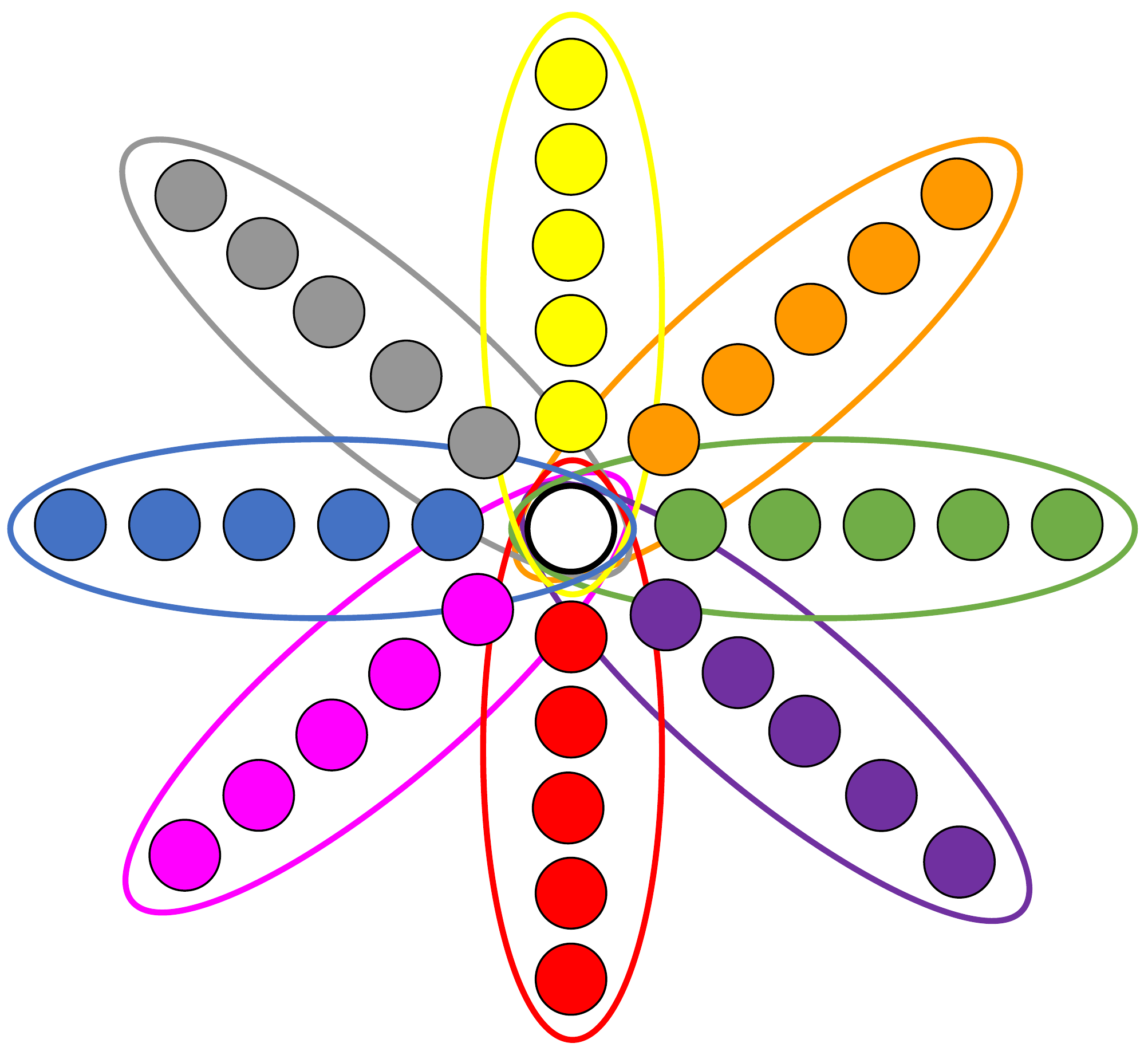} \label{fig: confs a}} \hspace{2cm} 
\subfloat[][]{\includegraphics[width=0.25\textwidth]{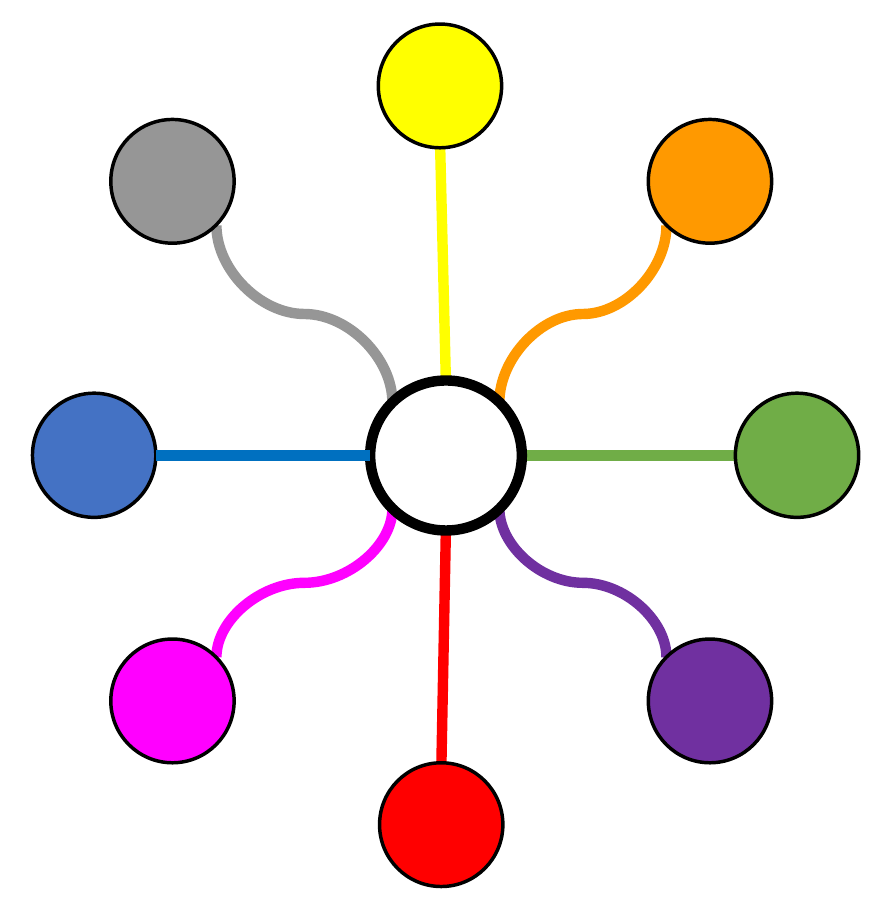} \label{fig: confs b}}
\captionsetup{justification=raggedright,format=plain,skip=4pt}%
\caption{(a) A spin-chain-star system composed by eight five-spin chains.
The central white circle represents the ancilla, while the ellipses embracing the spins of each chain and the ancilla represent the $N$-wise interactions.
The overlap of the eight ellipses is byproduct of the cartoon and has no physical meaning.
(b) The standard spin-star system unitarily equivalent to the system shown in panel (a).} \label{fig: confs}
\end{center}
\end{figure*}

In this work we consider particular spin-chain-star systems characterized by the peculiarity that the spins in each chain are all the same and interact through $N$-wise interactions among them and with the central spin, Fig.~\ref{fig: confs a}.
Such a kind of interactions can be implemented via quantum simulation apparatus based on either trapped ions \cite{Barreiro11,Muller11} or superconducting circuits made of transmon qubits \cite{Mezzacapo14}.
Moreover, these exotic couplings have been demonstrated to be exploitable to generate superposition states and then quantum correlations in large spin-chains under the experimental control of magnetic fields applied on the spin-system \cite{GLSM,GVdCVM}.

We analytically demonstrate that our spin-chain-star systems, under appropriate conditions, can be unitarily reduced to standard spin-1/2-star systems, Fig.~\ref{fig: confs b}.
This result implies the possibility of applying to such a large system some of the results already achieved for the standard spin-star systems, as, for example, the eigenspectrum, the eigenvalues and the quantum dynamics \cite{Villazon20}.
We show, indeed, that, by appropriately `translating' these results in the spin-chain-star language, it is possible to bring to light how to to generate entangled states of the different chains.
Further, we show as well how to produce different classes of entangled states depending on the specific topological configuration chosen for the spin-chain-star system.

The paper is organized as follows.
In Sec. \ref{SCSS} different types of $N$-wise spin-chain star models together with their unitarily equivalent standard spin-star models are introduced and their possible integrability is analysed.
The quantum dynamics and the possibility of generating different classes of entangled states (depending on the configuration) for an $XX$ spin-chain-star model are discussed in detail in Sec. \ref{QD}.
Finally, concluding remarks are reported in Sec. \ref{Conc}.

\section{Spin-chain-star systems} \label{SCSS}

In this section we propose a new class of spin-chain star systems characterized by the presence of only many-body interactions of maximum order.
We show that, by considering the many-body $N$-wise interactions for each chain, such systems are unitarily equivalent to standard spin-chain star systems, which are remarkably considered and studied for their applications in several fields \cite{Yung11,Sushkov14,He19,Arisoy21,Haddadi21, Ma10}.

\subsection{The $X$ Model}

Consider the following model:
\begin{equation} \label{Ham 2 chains}
    H = H_1 + H_2,
\end{equation}
with
\begin{subequations}
  \begin{align}
    H_1 &= {\hbar \omega_a \over 2} \hat{\sigma}_a^z + \gamma_1 \hat{\sigma}_a^x \otimes \left[ \bigotimes_i^{M_1} \hat{\sigma}_{1i}^x \right], \\
    H_2 &= {\hbar \omega_a \over 2} \hat{\sigma}_a^z + \gamma_2 \hat{\sigma}_a^x \otimes \left[ \ \bigotimes_j^{M_2} \hat{\sigma}_{2j}^x \right].
  \end{align}
\end{subequations}
The physical system described by this Hamiltonian model can be thought of as a central spin-1/2, to which we refer as ancilla, coupled to two spin-1/2 chains (the subscripts of the two terms in the Hamiltonian \eqref{Ham 2 chains} refer to the two different chains).
The coupling which characterizes the two spin-chains (including the ancilla) consists in $N$-wise interaction terms, that is, a type of interaction involving all the spins at one time.

It has been demonstrated \cite{GLSM} that the $M$-wise spin operator $\bigotimes_{i=1}^M \hat{\sigma}_{k}^x$ can be unitarily reduced as (see Appendix \ref{App})
\begin{equation} \label{Passage x}
    \bigotimes_{i=1}^M \hat{\sigma}_{i}^x ~ \rightarrow ~ \hat{\sigma}_1^x,
\end{equation}
where $\hat{\sigma}_1^x$ is intended to be a $2^M$-dimensional operator.
From a physical and mathematical point of view, it means that the $M$-spin chain effectively behaves as a single two-level system and then can be formally treated as a single qubit.
The origin of such a lucky circumstance can be traced back to the existence of a set of constants of motion which generate an equally numbered set of dynamically invariant two-dimensional Hilbert subspaces \cite{GLSM}.
This implies that, within each of these subspaces, the $M$-spin dynamics can be mapped into that of a single two-level system \cite{GLSM}.
A particularly interesting subdynamics is that characterized by the Hilbert space spanned by the two states $\ket{\downarrow}^{\otimes M}$ and $\ket{\uparrow}^{\otimes M}$ (with $\hat{\sigma}^z\ket{\downarrow}= -\ket{\downarrow}$ and $\hat{\sigma}^z\ket{\uparrow}= +\ket{\uparrow}$). 

Therefore, the two $M_k$-spin operators $\bigotimes_i^{M_1} \hat{\sigma}_{1i}^x$ and $\bigotimes_j^{M_2} \hat{\sigma}_{2j}^x$ in $H_1$ and $H_2$, can be unitarily transformed into $\hat{\sigma}_{11}^x$ and $\hat{\sigma}_{21}^x$, respectively.
In this way, the model in Eq. \eqref{Ham 2 chains}, after the appropriate unitary transformations, can be written as
\begin{equation}
    \tilde{H} = \hbar\omega_a \hat{\sigma}_a^z +
    \gamma_1 \hat{\sigma}_a^x \otimes \hat{\sigma}_{1}^x +
    \gamma_2 \hat{\sigma}_a^x \otimes \hat{\sigma}_{2}^x,
\end{equation}
where the second index of the spin operator in the last two terms, indicating the first spin of each chain, has been omitted.

Let us consider now a more general spin-chain system as the following one
\begin{equation} \label{X spin-chain-star}
    H_x = \hbar\omega_a \hat{\sigma}_a^z +
    \sum_k^N \gamma_k \hat{\sigma}_a^x \otimes \left[ \ \bigotimes_j^{M_k} \hat{\sigma}_{kj}^x \right].
\end{equation}
We can call such a physical system a spin-chain-star system since we can imagine the different $N$ chains to be disposed in a star-shaped configuration, each coupled to the same central spin.
It is easy to convince oneself that also this model, analogously to the two-chain case, can be transformed through unitary transformations into the following simpler spin model:
\begin{equation} \label{X spin-star}
    \tilde{H}_x = \hbar\omega_a \hat{\sigma}_a^z +
    \sum_k^N \gamma_k \hat{\sigma}_a^x \otimes \hat{\sigma}_{k}^x.
\end{equation}
This result shows that a star-shaped spin-chain system can be formally described and mathematically treated as a standard spin-star system, that is, a system consisting of $N$ mutually uncoupled spin-1/2's, each interacting with a unique central spin-1/2.
This possibility stems from the fact that each $N$-wise-interacting spin-chain can be effectively reduced to a single two-level system.

\subsection{The $XY$ Model}

It is interesting to point out that also the $N$-wise spin-operator $\bigotimes_{j=1}^M \hat{\sigma}_{j}^y$ can be unitarily reduced to \cite{GLSM} (see Appendix \ref{App})
\begin{equation} \label{Passage y}
    \bigotimes_{j=1}^M \hat{\sigma}_{j}^y ~ \rightarrow ~
    \Biggl[ (-1)^{M-1 \over 2} \gamma_{y} \prod_{j=1}^{(M-1)/2} \sigma_{2j+1}^{z} \Biggr] \hat{\sigma}_1^y,
\end{equation}
provided that the number $M$ of spins is odd \cite{GLSM}, as shown in Appendix \ref{App}.
The operator realizing such a transformation is the same accomplishing that in Eq. \eqref{Passage x}.
In the above expressions $\sigma_{2j+1}^{z}$ are constants of motion and their possible values can be +1 and -1.
The specific values assigned to these integrals of motion identify a precise dynamically invariant subspace \cite{GLSM}.
The Hilbert subspace spanned by $\ket{\downarrow}^{\otimes M}$ and $\ket{\uparrow}^{\otimes M}$, for example, is characterized by all the constants of motion equal to 1, namely $\sigma_{2j+1}^{z}=1, ~ \forall ~ j$.
Therefore, the following $XY$ spin-chain-star system
\begin{equation} \label{XY spin-chain-star}
    H_{xy} = \hbar\omega_a \hat{\sigma}_a^z +
    \sum_k^N \gamma_k^x \hat{\sigma}_a^x \otimes \left[ \bigotimes_j^{M_k} \hat{\sigma}_{kj}^x \right]+
    \sum_k^N \gamma_k^y \hat{\sigma}_a^y \otimes \left[ \bigotimes_j^{M_k} \hat{\sigma}_{kj}^y \right],
\end{equation}
can be mapped into the standard $XY$ spin-star system, which reads
\begin{equation} \label{XY spin-star}
    \tilde{H}_{xy} = \hbar\omega_a \hat{\sigma}_a^z +
    \sum_k^N \gamma_k^x \hat{\sigma}_a^x \otimes \hat{\sigma}_{k}^x +
    \sum_k^N \gamma_k^y \hat{\sigma}_a^y \otimes \hat{\sigma}_{k}^y,
\end{equation}
if the number $M_k$ of spins in the $k$-th chain is chosen so that $(M_k-1)/2$ is even, and if all the chains coupled to the central ancilla are initially prepared in either $\ket{\downarrow}^{\otimes M_k}$ or $\ket{\uparrow}^{\otimes M_k}$ or any arbitrary superposition of these two states (such as a GHZ-like state).

\subsection{The $XYZ$ Model}

It is possible to demonstrate that the same set of unitary operators realizing the transformations in Eqs. \eqref{Passage x} and \eqref{Passage y} realizes also the following transformation \cite{GLSM} (see Appendix \ref{App})
\begin{equation} \label{Passage z}
    \bigotimes_{l=1}^M \hat{\sigma}_{l}^z ~ \rightarrow ~
    \Biggl[ \gamma_{z} \prod_{l=1}^{(M-1)/2} \sigma_{2l+1}^{z} \Biggr] \hat{\sigma}_1^z,
\end{equation}
where $M$ is odd.

Also in this case, if the involved two-level subdynamics of each chain is that characterized by the two states $\ket{\downarrow}^{\otimes M_k}$ and $\ket{\uparrow}^{\otimes M_k}$, then the $XYZ$ spin-chain-star system
\begin{equation} \label{XYZ spin-chain-star}
  \begin{aligned}
    H_{xyz} = & \hbar\omega_a \hat{\sigma}_a^z +
    \sum_k^N \gamma_k^x \hat{\sigma}_a^x \otimes \left[ \bigotimes_j^{M_k} \hat{\sigma}_{kj}^x \right]+ \\
    & \sum_k^N \gamma_k^y \hat{\sigma}_a^y \otimes \left[ \bigotimes_j^{M_k} \hat{\sigma}_{kj}^y \right]+
    \sum_k^N \gamma_k^z \hat{\sigma}_a^z \otimes \left[ \bigotimes_j^{M_k} \hat{\sigma}_{kj}^z \right],
  \end{aligned}
\end{equation}
is unitarily equivalent to the standard $XYZ$ spin-star model
\begin{equation} \label{XYZ spin-star}
  \begin{aligned}
    \tilde{H}_{xyz} = & \hbar\omega_a \hat{\sigma}_a^z +
    \sum_k^N \gamma_k^x \hat{\sigma}_a^x \otimes \hat{\sigma}_{k}^x + \\
    & \sum_k^N \gamma_k^y \hat{\sigma}_a^y \otimes \hat{\sigma}_{k}^y +
    \sum_k^N \gamma_k^z \hat{\sigma}_a^z \otimes \hat{\sigma}_{k}^z.
  \end{aligned}
\end{equation}
A qualitative representation of a star-shaped system composed by eight five-spin chains is shown in Fig.~\ref{fig: confs a}.
Its unitarily equivalent standard spin-star system is shown in Fig.~\ref{fig: confs b}.

Finally, it is worth pointing out that the effective mathematical description, basing on the unitary transformation procedure, is not affected by a possible time-dependence of the Hamiltonian parameters.
This property stems from the fact that the unitary operators, which transform the Hamiltonians, does not depend on the Hamiltonian parameters and, more in general, on time.

\subsection{In presence of fields}

In this subsection we see that our analysis and then the unitary reduction of a spin-chain-star model to a standard spin-star model keeps its validity also when fields applied to the entire chains are considered.
Let us suppose each entire chain in the system to be subject to a uniform magnetic field.
The models in Eqs. \eqref{X spin-chain-star}, \eqref{XY spin-chain-star} and \eqref{XYZ spin-chain-star} are then enriched of the term
\begin{equation} \label{Fields on chains}
    \sum_k^N \hbar \omega_0^k \sum_j^{M_k} \hat{\sigma}_{kj}^z.
\end{equation}

It is possible to convince oneself \cite{GLSM} that the unitary transformations acting as expressed in Eqs. \eqref{Passage x}, \eqref{Passage y} and \eqref{Passage z} convert the operators in Eq. \eqref{Fields on chains} into (see Appendix \ref{App})
\begin{equation}
    \sum_k^N \left[ 1 + \sum_{j=2}^{M_k} \prod_{i=2}^j \sigma_{kj}^z \right] \hbar \omega_0^k \hat{\sigma}_{k1}^z,
\end{equation}
where $\sigma_{kj}^z$ are constants of motions as before.
Within the subspace (for each chain) we are interested in, that is the one spanned by the two states $\ket{\downarrow}^{\otimes M_k}$ and $\ket{\uparrow}^{\otimes M_k}$, the integrals of motions are all equal to 1.
We can then write the following effective two-level operators
\begin{equation}
    \sum_k^N M_k ~ \hbar \omega_0^k ~ \hat{\sigma}_{k}^z,
\end{equation}
each of which accounts for the magnetic field applied on a chain, whose dynamics is equivalent to that of a single spin-qubit system.

Therefore, in this physical scenario, the unitarily transformed effective models in Eqs. \eqref{X spin-star}, \eqref{XY spin-star} and \eqref{XYZ spin-star} are modified by simply introducing such terms in the Hamiltonians.
It is important to underline that the introduction of the fields uniformly acting upon each chain does not alter the symmetry properties of the original Hamiltonians.
In this way, the possibility to perform the same unitary operations on the Hamiltonian interaction terms results to be not affected.

As a final remark, we wish to emphasize the importance of the symmetry properties possessed by the Hamiltonian(s).
For the case analysed in this paper, the study of the Hamiltonian symmetries is fundamental for the exact solution of the dynamical problem.
More in general, the symmetries, besides allowing to analytically treat some models, have profound physical implications.
Indeed, the disclosure of symmetry-protected (sub-)dynamics in different physical systems \cite{GMIV, GMMM} can lead to discover physical effects which turn out to be useful and applicable in different fields such as quantum metrology \cite{Yoshinaga21, Hatomura22}.

\subsection{Integrability} \label{Int}
 
It is worth pointing out that the fully isotropic $XXX$ spin-star model is integrable; precisely, it belongs to the class of $XXX$ Richardson-Gaudin integrable
models \cite{Gaudin14,Villazon20}.
The condition of integrability stems from the existence of an appropriate set of
integrals of motion, allowing to obtain all eigenstates and related eigenvalues through the use of Bethe ansatz techniques\cite{Villazon20}.
This circumstance, joined with the fact that the $XXX$ spin-star model well describes systems with spherical symmetry (such as quantum dots in semiconductors with s-type conduction bands \cite{Hanson07}), has spurred several studies focused on the equilibrium and dynamical properties of such a model \cite{Claeys18,Faribault13}.

Very recently, it has been demonstrated that also the $XX$ model is integrable \cite{Villazon20}.
This model naturally emerges in resonant dipolar spin systems in rotating frames \cite{Fernandez18,Ding14} and its eigenstates are divided into two classes: dark and bright states.
The former are product states of the ancilla and of all the other environmental spins, so that the central spin is thus disentangled from the spin-bath.
The latter can be written as a combination of dark states and then exhibit entanglement between the ancilla and the other spins \cite{Villazon20}.

On this basis we therefore claim that, when an $XX$ spin-chain star model can be unitarily reduced to a standard spin-star one, we can derive the exact expressions of the eigenvalues and eigenvectors of the spin-chain-star system.
It is important to stress that, in this case, fully integrability cannot be invoked since the spin-chain-star model is exactly solvable only within the specific subspaces where the mapping to an integrable spin-star model is possible.
As previously said, one of these subspaces is that spanned by the pair of states $\{ \ket{\uparrow}^{\otimes M}, \ket{\downarrow}^{\otimes M} \}$.
Within other subspaces, instead, although the reduction to standard spin-star models is always possible, the effective unitarily equivalent models present inhomogeneities ($XYZ$) which affect the integrability.

\section{Exact solution for $XX$ spin-chain-star systems} \label{QD}

\subsection{$W$-like and $GHZ$-like states of spin-chains}

In this section we specialize the $XX$ spin-chain-star system considered so far by setting $\gamma_k^x=\gamma_k^y, ~ \forall~k$ in Eq. \eqref{XY spin-chain-star}, namely
\begin{equation} \label{XX spin-chain-star}
    H_{xx} = \hbar\omega_a \hat{\sigma}_a^z + \sum_k^N \gamma_k \left\{ \hat{\sigma}_a^x \otimes \left[ \bigotimes_j^{M} \hat{\sigma}_{kj}^x \right]+
    \hat{\sigma}_a^y \otimes \left[ \bigotimes_j^{M} \hat{\sigma}_{kj}^y \right] \right\}.
\end{equation}
Moreover, we suppose a number $N$ of spin-1/2-chains, each consisting of $M$ spin-qubits and satisfying the constraint that $(M-1)/2$ is an even number.
Through the appropriate unitary transformations previously discussed, we get thus an effective standard $XX$ spin-star system (Eq. \eqref{XY spin-star} with  $\gamma_k^x=\gamma_k^y, ~ \forall~k$)
\begin{equation} \label{XX spin-star}
    \tilde{H}_{xx} = \hbar\omega_a \hat{\sigma}_a^z +
    \sum_k^N \gamma_k \left[ \hat{\sigma}_a^x \otimes \hat{\sigma}_{k}^x +
    \hat{\sigma}_a^y \otimes \hat{\sigma}_{k}^y \right].
\end{equation}

We consider all the $N$ spin-chains initialized in the $M$-spin state $\ket{\downarrow}^{\otimes M}$.
As said in the previous section, the symmetries of the Hamiltonian lead to a two-dimensional dynamically invariant subspace spanned by $\ket{\downarrow}^{\otimes M}$ and $\ket{\uparrow}^{\otimes M}$.
It means that the $k$-th chain ($k=1 \dots N$) can be effectively represented in terms of dynamical variables $\sigma_k^x, ~ \sigma_k^y, ~ \sigma_k^z$ of a fictitious qubit.
The following mapping (valid for each chain)
\begin{equation} \label{Mapping}
    \ket{\downarrow}^{\otimes M} \Longleftrightarrow \ket{-},
    \qquad
    \ket{\uparrow}^{\otimes M} \Longleftrightarrow \ket{+}.
\end{equation}
(with $\hat{\sigma}^z\ket{\pm}= \pm\ket{\pm}$) enables to fix unambiguously the initial state of the fictitious standard spin-star system.
We suppose the ancilla and the effective standard spin-star system initially prepared in the following state
\begin{equation} \label{Initial state}
    \ket{\psi(0)} = \ket{\uparrow_a} \ket{-}^{\otimes N},
\end{equation}
which in terms of spin-chain states is written as
\begin{equation} \label{Initial state new}
    \ket{\psi(0)} = \ket{\uparrow_a} \ket{\downarrow_1}^{\otimes M} \dots \ket{\downarrow_k}^{\otimes M} \dots \ket{\downarrow_N}^{\otimes M}.
\end{equation}

It is known that the exact time evolution of the initial condition taken into account for the $XX$ spin-star system is \cite{Jivulescu09,Ferraro08}
\begin{equation} \label{Evolved state}
    \ket{\psi(t)} = \alpha(t) \ket{\uparrow_a} \ket{-}^{\otimes N} + \ket{\downarrow_a} \sum_{k}^N \beta_k(t) \ket{-_1 \dots +_k \dots -_N},
\end{equation}
with
\begin{subequations} \label{alpha beta}
  \begin{align}
    \alpha(t) &= \cos(\omega~t) + i {\omega_a \over \omega} \sin(\omega~t) \label{alpha}, \\
    \beta_k(t) &= -i {\gamma_k/\hbar \over \omega} \sin(\omega~t), \label{beta}
  \end{align}
\end{subequations}
where
\begin{equation} \label{omega}
    \omega=\sqrt{\sum_k (\gamma_k/\hbar)^2 + \omega_a^2}.
\end{equation}

We underline that, for the initial condition under scrutiny, only one effective frequency, namely $\omega$, characterizes the time evolution of the system.
Thus, the $XX$ spin-star system comes back to its initial condition with a period $T = 2\pi/\omega$ and behaves as if its dynamics were governed by an effective Hamiltonian describing $N$ different spins homogeneously coupled to the central one, that is with the same effective coupling constant.
Precisely, such an effective model can be obtained by substituting in Eq. \eqref{XY spin-chain-star} $\gamma_k^x=\gamma_k^y=\gamma_{eff}$ with $\gamma_{eff} = \hbar \omega$, which is independent of $k$.

\begin{figure}[htp] 
\begin{center}
{\includegraphics[width=0.45\textwidth]{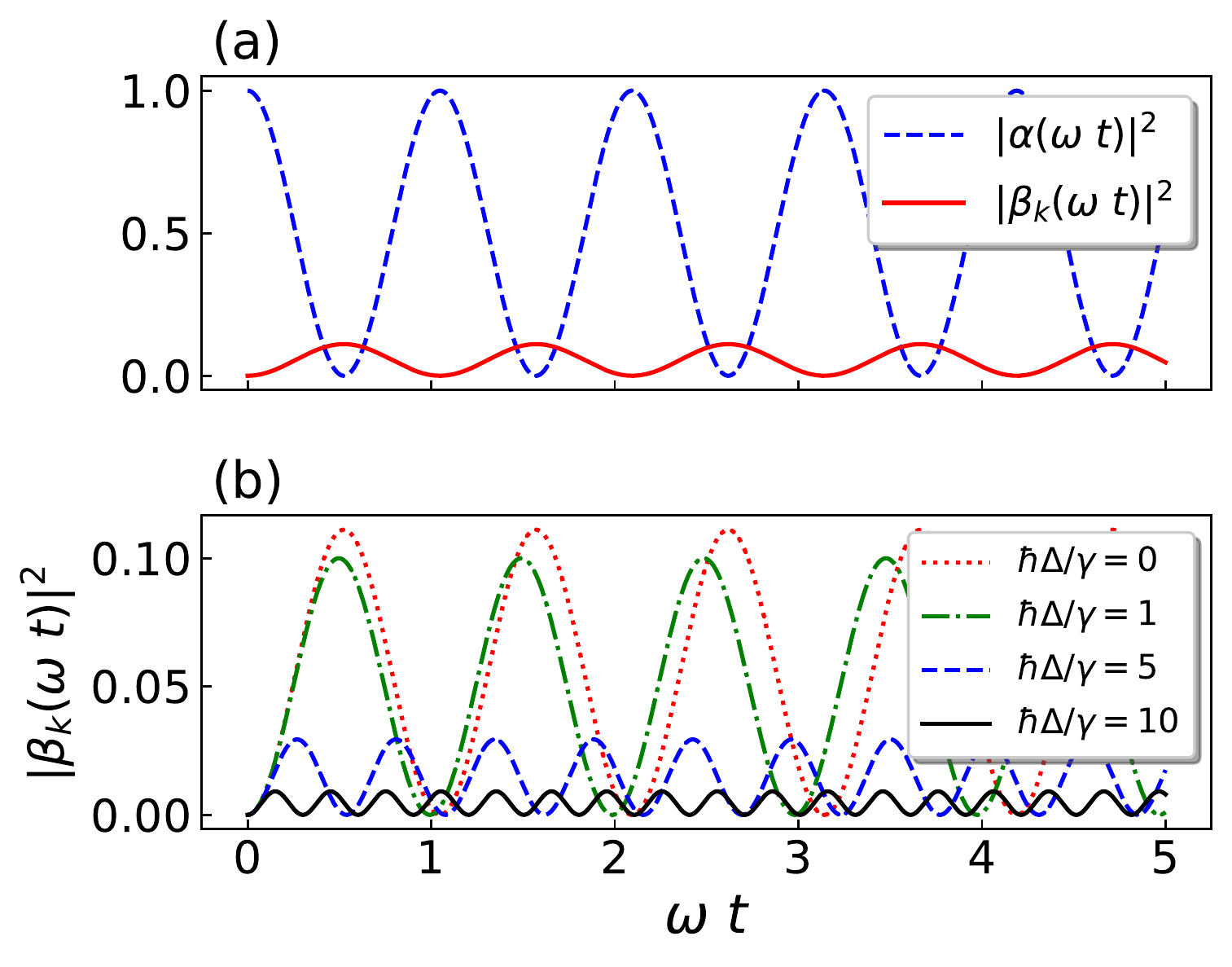}}
\captionsetup{justification=raggedright,format=plain,skip=4pt}%
\caption{(a) Time dependencies of $|\alpha(t)|^2$ (blue dashed line) and $|\beta_k(t)|^2  = |\beta(t)|^2, ~ \forall ~ k$ (red solid line) [see Eq. \eqref{alpha beta}] in case of $N=9$ number of $M$-spin chains, for $\gamma_k=\gamma, ~ \forall ~ k$ and $\omega_a=0$.
(b) Effects of the detuning on the time dependence of $|\beta_k(t)|^2  = |\beta(t)|^2, ~ \forall ~ k$, in case of $N=9$ and $\gamma_k=\gamma, ~ \forall ~ k$; different values of the ratio $\hbar \Delta / \gamma$ are considered: 0 (red dotted line), 1 (green dot-dashed line), 5 (blue dashed line) and 10 (black solid line).} \label{fig: alpha-beta}
\end{center}
\end{figure}

Moreover, it is interesting to note that, for $\gamma_k = \gamma, ~ \forall ~ k$ and $\omega_a=0$, when $t=n \pi / \omega$ (so that $\alpha(t)=0$), we get the state
\begin{equation} \label{W-states}
    \ket{W} = \ket{\downarrow_a} \left[ {1 \over \sqrt{N}} \sum_{k}^N \ket{-_1 \dots +_k \dots -_N} \right],
\end{equation}
up to a global factor $\exp\{-i \pi/2\}$.
For this scenario, the time behaviours of the probabilities $|\alpha(t)|^2$ and $|\beta_k(t)|^2 = |\beta(t)|^2, ~ \forall ~ k$, are shown in Fig. \ref{fig: alpha-beta}(a) in the case of $N=9$.

We highlight that the $N$-two-level system results in a $W$-like state and that, therefore, each of these $N$ involved two-level systems represents one of the $N$ $M$-spin chains.
Thus, the ancilla-mediated (quantum) correlations which arise between the effective $N$ spin-1/2's can be interpreted as correlations get established between the $N$ spin-chains.
It means that such a $W$-state is a `macro-state' consisting in a superposition of states involving all the $M$-spin-chains in the systems.
We can then write the $W$-like state in terms of the spin-chains as
\begin{equation} \label{W-states spin-chain}
    \ket{W} = \ket{\downarrow_a} \left[ {1 \over \sqrt{N}} \sum_{k}^N \ket{\downarrow_1}^{\otimes M} \dots \ket{\uparrow_k}^{\otimes M} \dots \ket{\downarrow_N}^{\otimes M} \right].
\end{equation}
Therefore, we claim that the $XX$ spin-chain-star model [Eq. \eqref{XX spin-chain-star}] allows for the generation of `macro' $W$-like states of the chains and then for the creation of a large-scale entanglement between all the subsystems in the spin-chain-star physical scenario. 

\subsection{Effects of Detuning}

Let us suppose to apply a uniform  magnetic field to all the identical $M$-spin-chains of the system, namely
\begin{equation} \label{Fields on chains det}
     {\hbar \omega_0 \over M} \sum_k^N \sum_j^{M} \hat{\sigma}_{kj}^z.
\end{equation}
In this case the unitarily equivalent spin-star model reads
\begin{equation} \label{XX spin-star fields}
    \tilde{H}_{xx}' = \hbar\omega_a \hat{\sigma}_a^z + 
    \hbar \omega_0 \sum_k^N \hat{\sigma}_{k}^z +
    \sum_k^N \gamma_k \left[ \hat{\sigma}_a^x \otimes \hat{\sigma}_{k}^x +
    \hat{\sigma}_a^y \otimes \hat{\sigma}_{k}^y \right].
\end{equation}

It is easy to check that, for such a physical scenario, the expression of the evolved state in Eq. \eqref{Evolved state}, obtained when the spin-chain-star system is initially prepared in the state \eqref{Initial state}, remains formally identical, as well as that of $\beta_k(t)$ in Eq. \eqref{beta} \cite{Jivulescu09,Ferraro08}.
The only slight variation is found in the mathematical expressions of $\alpha(t)$ and $\omega$ which become now respectively \cite{Jivulescu09,Ferraro08}
\begin{subequations} 
  \begin{align}
    \alpha(t) &= \cos(\omega~t) - i ~ {\Delta \over \omega} \sin(\omega~t) \label{alpha det}, \\
    \omega &= \sqrt{\sum_k (\gamma_k/\hbar)^2 + \Delta^2},   
  \end{align}
\end{subequations}
where the detuning is defined as follows:
\begin{equation} \label{omega det}
    \Delta = \omega_0 - \omega_a.
\end{equation}
The effects of the detuning on the probabilities $|\beta_k(t)|^2$ are shown in Fig. \ref{fig: alpha-beta}(b), for $\gamma_k=\gamma, ~ \forall ~ k$.
We see, as expected, that the greater is the detuning, the lower are the probabilities $|\beta_k(t)|^2$ (and then the higher the complementary probability $|\alpha(t)|^2$), meaning that the system tends to remain in its initial condition for high values of the detuning.

This result, thus, demonstrates also that, although a non-vanishing field $\omega_a$ is present on the ancilla spin, it is still possible generating the $W$-state in Eq. \eqref{W-states}, provided that a further magnetic field of magnitude $\omega_a/M$ is uniformly applied to all the $M$-spin chains (so that $\Delta=0$).

\subsection{Entanglement}

In this section we investigate the possible quantum correlations emerging in the spin-chain-star system.
To this end, we consider the case of $N=2$ chains, each made of $M$ spin-qubits, and $\gamma_k = \gamma, ~ \forall ~ k$. 
It is interesting to note that in this case, through the procedure previously exposed and for $\Delta=0$, if we get $-1$ by measuring the ancilla dynamical variable $\hat{\sigma}_a^z$ at $t= \pi/\omega$, Eq. \eqref{Evolved state} foresees that the resulting spin-state of the two chains has the following form
\begin{equation} \label{GHZ 2 chains}
    \ket{GHZ} = {\ket{\uparrow}^{\otimes M} \ket{\downarrow}^{\otimes M} + \ket{\downarrow}^{\otimes M} \ket{\uparrow}^{\otimes M} \over \sqrt{2}}.
\end{equation}
Since in this state the concurrence \cite{Wootters} between two generic spin-1/2’s (in the same or different chains) vanishes and the measure of the collective $z$-component of the whole spin-chain can give only the two values $\pm N$, it is legitimate to call this multi-spin state a $GHZ$-like state. 
The legitimacy of such a denomination of the state \eqref{GHZ 2 chains} can be convincingly strengthened simply observing that, exploiting our mapping \eqref{Mapping}, the same state \eqref{GHZ 2 chains} can be written as the Bell state
\begin{equation}
    \ket{GHZ} \rightarrow \ket{\Psi^+} = {\ket{+} \ket{-} + \ket{-} \ket{+} \over \sqrt{2}},
\end{equation}
which is characterized by the maximum level of concurrence ($C=1$).
We stress that, as a result of our analysis, the entanglement should be intended as the signature of the existence of quantum correlations between the spin-chains. 
We wish to emphasize the added value of the representation in terms of fictitious two-level systems, since it provides information about quantum correlations get established between the spin-chains.
According to this interpretation, it becomes relevant to calculate the time dependence of the concurrence exhibited by two chains in the state \eqref{Evolved state}:
\begin{equation}
  \begin{aligned}
    \ket{\psi(t)} =& \alpha(t) \ket{\uparrow_a} \ket{-} \ket{-} \\
    &+\ket{\downarrow_a} \left[ \beta_1 (t) \ket{+} \ket{-} + \beta_2(t) \ket{-} \ket{+}  \right] = \\
    =&\alpha(t) \ket{\uparrow_a} \ket{\downarrow}^{\otimes M} \ket{\downarrow}^{\otimes M} \\
    &+\ket{\downarrow_a} \left[ \beta_1 (t) \ket{\uparrow}^{\otimes M} \ket{\downarrow}^{\otimes M} + \beta_2(t) \ket{\downarrow}^{\otimes M} \ket{\uparrow}^{\otimes M}  \right].
  \end{aligned}
\end{equation}
It is possible to verify that the concurrence for the density matrix of the two effective spin-1/2's is
\begin{equation}
    C(t) = 2 |\beta_1(t)| |\beta_2(t)|.
\end{equation}
We are therefore able to write the time-dependence of the entanglement emerging between the two chains.
The latter results to be maximum for $\beta_1(t) = \beta_2(t) = 1/\sqrt{2}$, corresponding to the $GHZ$-like state previously examined.

Analogously, in the case of $N=3$ $M$-spin chains, by measuring $\sigma_a^z=1$ at $t= \pi/\omega$, the state reached by the spin system is
\begin{equation}
  \begin{aligned}
    &\ket{W} =  { \ket{+--} + \ket{-+-} + \ket{--+} \over \sqrt{3}} = \\
    &{\ket{\uparrow}^{\otimes M} \ket{\downarrow}^{\otimes M} \ket{\downarrow}^{\otimes M} + \ket{\downarrow}^{\otimes M} \ket{\uparrow}^{\otimes M} \ket{\downarrow}^{\otimes M} + \ket{\downarrow}^{\otimes M} \ket{\downarrow}^{\otimes M} \ket{\uparrow}^{\otimes M} \over \sqrt{3}},
  \end{aligned}
\end{equation}
which can be interpreted as a maximally entangled $W$-like state of the three $M$-spin chains.
Also in this case we may infer the entanglement between the three spin chains with the help of the effective description involving three interacting two-level systems as well.
It is easy to verify that, for the state \eqref{Evolved state} in the case of $N=3$, each pair $i$-$j$ of chains exhibits a non-vanishing concurrence equal to $2|\beta_i||\beta_j|$.
Each pair of two generic true spin-1/2's is, instead, disentangled as in the case of two chains.

This result can be extended to the case of $N$ spin-chains.
In this instance, it is possible to check, indeed, that the entanglement between two generic true spins vanishes, while the concurrence for a generic pair of chains $i$-$j$ results to be $2|\beta_i||\beta_j|$. 

This means that the spin-chain-star system here discussed, besides generating quantum correlations in a large spin-system, is suitable to generate different types of entangled states of the chains assumed in the model under scrutiny.
The origin of such differences can be traced back to specific topological and structural properties  of the spin-chain-star system, as for example the number of chains and the number of spins per chain.

\section{Conclusions} \label{Conc}

In this work we have considered a special class of spin-chain-star systems.
Precisely, we have focused our attention on spin chains characterized by many body $N$-wise interactions, that is interaction terms involving all the spins in a chain at once.
We have taken into account several types of spin-chain-star models including different types of $N$-wise interactions as well as the presence of local magnetic fields on both the ancilla and the spins in the chains.

We have shown that each model we have considered can be analytically treated through unitary transformations and that each chain, for specific initial conditions, effectively behaves and can be thought of as a two-level system.
This implies that a spin-chain-star system belonging to the class under scrutiny is unitarily equivalent to a standard spin-star system.
Therefore, we can exploit the knowledge and the results obtained for the standard spin-star models and interpret them in terms of multiple-chain states.
For example, we have demonstrated that, by starting from a disentangled state of the spin-chain-star system, our model and scheme allow for the generation of a `macro'-entangled state of all the spin chains which form the system.
Therefore, we can speak of quantum correlations arising between the actual spins and between the spin chains globally described as two-level systems.
In case of $N=2$ and $N=3$ spin chains, thanks to the mapping into a spin-1/2-star model, we can affirm that the system evolves and reaches a maximally entangled superposition at appropriate instants of time.
Moreover, we are also able to quantify the entanglement get established between two spin chains through the calculation of the concurrence (since the two chains effectively behave as two spin-qubits).

Finally, we have analysed the effects on the probability related to the generation of such a state stemming from the presence of magnetic fields acting on the ancilla and (homogeneously) on the chains.
In this way, we have found the optimal experimental work condition necessary to get a macro-entangled state.
Our result, thus, paves the way to the possibility of generating large-scale entanglement in spin systems made of several spins.

Further investigations could concern quantum oscillator baths(s) interacting with the spin chains and/or the ancilla.
In this case, likely, a numerical approach to solve the dynamics is necessary.
Several approaches could be used to deal with this scenario, from the standard Lindblad theory \cite{Gorini76, Lindblad76} to the partial Wigner transform \cite{Kapral99, kapral01, Sergi03, Sergi05, Sergi07, SGHM, SHGM} and the non-Hermitian formalism \cite{Feshbach58, Bender98, Mostafazadeh10, Rotter15, Sergi13, Sergi15, Brody12, GdCKM, GdCNM2, GMSVF}.

\section*{Acknowledgements}

RG and DV acknowledge financial support from the PRIN Project PRJ-0232 - Impact of Climate Change on the biogeochemistry of Contaminants in the Mediterranean sea (ICCC).

\section*{Appendices}

\appendix

\section{Symmetry-based unitary transformations} \label{App}

Let us consider the following $N$-spin model
\begin{equation}\label{HNspinMessinaSM}
H=\gamma_{x}\bigotimes_{k=1}^{N}\hat{\sigma}_{k}^{x}+
\gamma_{y}\bigotimes_{k=1}^{N}\hat{\sigma}_{k}^{y}+
\gamma_{z}\bigotimes_{k=1}^{N}\hat{\sigma}_{k}^{z},
\end{equation}
where $\hat{\sigma}^x$, $\hat{\sigma}^y$ and $\hat{\sigma}^z$ are the standard Pauli matrices.
The $N$ distinguishable spins are coupled only through $N$-wise interaction terms, that is interaction terms which involve all the $N$-spins at once.

To exactly diagonalize the model, it is useful to begin with the easiest case of two interacting spin-1/2's \cite{GMN, GVM1, GIMGM, GNMV}:
\begin{equation}\label{H2SpinMessinaSM}
H_2= 
\gamma_{x}\hat{\sigma}_{1}^{x}\hat{\sigma}_{2}^{x}+\gamma_{y}\hat{\sigma}_{1}^{y}\hat{\sigma}_{2}^{y}+\gamma_{z}\hat{\sigma}_{1}^{z}\hat{\sigma}_{2}^{z},
\end{equation}
for which $\hat{\sigma}_{1}^{z}\hat{\sigma}_{2}^{z}$ is an integral of motion, $[H_2,\hat{\sigma}_{1}^{z}\hat{\sigma}_{2}^{z}]=0$.
It is possible to verify that the following unitary and Hermitian operator ($\mathbb{1}$ is the identity operator in the four dimensional Hilbert subspace)
\begin{equation}\label{OperatoreUH2spinMessinaSM}
\mathbb{T}_{12}=\dfrac{1}{2}\left[\mathbb{1}+\hat{\sigma}_{1}^{z}+\hat{\sigma}_{2}^{x}-\hat{\sigma}_{1}^{z}\hat{\sigma}_{2}^{x}\right],
\end{equation}
transforms $H_2$ into
\begin{equation}\label{Htilde2SpinMessinaTransfSM}
\mathbb{T}_{12}^{\dagger}H_2\mathbb{T}_{12}=\tilde{H}_{2}=
\gamma_{x}\hat{\sigma}_{1}^{x}-\gamma_{y}\hat{\sigma}_{2}^{z}\hat{\sigma}_{1}^{x}+\gamma_{z}\hat{\sigma}_{2}^{z}.
\end{equation}
Since $\hat{\sigma}_{2}^{z}$ is a constant of motion for $\tilde{H}$, it can be treated as a parameter ($=\pm 1$), rewriting
\begin{equation}\label{Htilde2SpinMessinaTransfParamSM}
\tilde{H}_{\sigma_{2}^{z}}=
\left(\gamma_{x}-\gamma_{y}\sigma_{2}^{z}\right)\hat{\sigma}_{1}^{x}+\gamma_{z}\sigma_{2}^{z}.
\end{equation}
The existence of such a symmetry (giving rise to the constant of motion) implies the existence of two dynamically invariant Hilbert subspaces, each of which related to an eigenvalue of $\sigma_{2}^{z}$.
Each single-spin-1/2 Hamiltonian, obtainable by assigning a value to $\sigma_{2}^{z}$, governs the two-spin dynamics in one of the two dynamically invariant Hilbert subspaces.
Therefore, in this way, we have reduced the two-interacting-spin problem to two independent single-spin-1/2 problems.

Let us consider now the analogous three-spin model
\begin{equation}\label{H3spinMessinaSM}
\begin{aligned}
H_3 = \gamma_{x}\hat{\sigma}_{1}^{x}\hat{\sigma}_{2}^{x}\hat{\sigma}_{3}^{x}+
\gamma_{y}\hat{\sigma}_{1}^{y}\hat{\sigma}_{2}^{y}\hat{\sigma}_{3}^{y}+
\gamma_{z}\hat{\sigma}_{1}^{z}\hat{\sigma}_{2}^{z}\hat{\sigma}_{3}^{z}.
\end{aligned}
\end{equation}
This time the new constant of motion $\hat{\sigma}_{2}^{z}\hat{\sigma}_{3}^{z}$ appears.
By applying the analogous procedure used for the two-spin case, we get the following new Hamiltonian
\begin{equation}\label{Htilde3spinMessinaSM}
\begin{aligned}
\mathbb{T}_{23}^{\dagger}H_3\mathbb{T}_{23}=&\tilde{H}_{3}=
\gamma_{x}\hat{\sigma}_{1}^{x}\hat{\sigma}_{2}^{x}-\gamma_{y}\sigma_{3}^{z}\hat{\sigma}_{1}^{y}\hat{\sigma}_{2}^{x}+\gamma_{z}\sigma_{3}^{z}\hat{\sigma}_{1}^{z}.
\end{aligned}
\end{equation}
In this case, the unitary transformation involves the second and the third spin, and the operator accomplishing such a transformation, inspired by $\mathbb{T}_{12}$, reads
\begin{equation}\label{OperatoreTrasfUH3spinMessinaSM}
\mathbb{T}_{23}=\dfrac{1}{2}\left[\mathbb{1}+\hat{\sigma}_{2}^{z}+\hat{\sigma}_{3}^{x}-\hat{\sigma}_{2}^{z}\hat{\sigma}_{3}^{x}\right].
\end{equation}
Since $\hat{\sigma}_{1}^{z}\hat{\sigma}_{2}^{z}$ is a constant of motion for $\tilde{H}_3$, we can exploit the operator written in Eq. \eqref{OperatoreUH2spinMessinaSM} and apply one more time the same procedure to $\tilde{H}_{3}$.
So, we get
\begin{equation}\label{Hbitilde3spinMessinaSM}
\begin{aligned}
\mathbb{T^{\dagger}}_{12}\tilde{H}_{3}\mathbb{T}_{12}&=
\mathbb{T^{\dagger}}_{123}H_{3}\mathbb{T}_{123}=
\tilde{\tilde{H}}_{3} \\
&=\gamma_{x}\hat{\sigma}_{1}^{x}-\gamma_{y}\sigma_{3}^{z}\hat{\sigma}_{1}^{y}+\gamma_{z}\sigma_{3}^{z}\hat{\sigma}_{1}^{z},
\end{aligned}
\end{equation}
where we put $\mathbb{T}_{123}=\mathbb{T}_{23}\mathbb{T}_{12}$.
In this case we have four dynamically invariant subspaces related to the four pairs of eigenvalues of the two constants of motion $\hat{\sigma}_{1}^{z}\hat{\sigma}_{2}^{z}$ and $\hat{\sigma}_{2}^{z}\hat{\sigma}_{3}^{z}$.
So, we have four two-level Hamiltonians governing the three-spin dynamics in each subspace.
Therefore, we have reduced the initial dynamical problem of three interacting spins to four independent single-spin-1/2 dynamical problems.

Basing on this last result, it is easy to understand that we can apply the same sequence of unitary transformations also in the case of $N$ spins.
More precisely, we can iterate the transformation procedure until the initial $N$-spin Hamiltonian is completely reduced to a set of $2^{N-1}$ two-level Hamiltonians.
Each of these effective single-spin-1/2 Hamiltonians governs the $N$-spin dynamics within an invariant subspace identified by the specific values of the $2^{N-1}$ constants of motion.
The total unitary operator accomplishing the chain of unitary transformations can be written as
\begin{equation}\label{OperatoreDiagonalizzanteHNspinSM}
\mathbb{T}
=\dfrac{1}{2^{N-1}}\prod_{k=0}^{N-2}\left[\mathbb{1}+\hat{\sigma}_{N-(k-1)}^{z}+\hat{\sigma}_{N-k}^{x}-\hat{\sigma}_{N-(k+1)}^{z}\hat{\sigma}_{N-k}^{x}\right],
\end{equation}
where each piece of the product acts on a `spin-triplet', say $[i,j,k]$.
As shown in the two- and three-spin cases, each three-spin transformation leaves the Hamiltonian dependent on the dynamical variables of the first spin of the triplet ($i$-th spin) and on those of all other spins not affected by the transformation.
The spins $j$ and $k$ appears only with the $z$-component, i.e. $\hat{\sigma}_{j}^{z}$ and $\hat{\sigma}_{k}^{z}$ which are constants of motion.
They can be therefore treated as parameters and substituted with their eigenvalues in the expression of the transformed Hamiltonian.

The effects on the Hamiltonian after the spin-triplet transform are:
\begin{itemize}
\item
a -1 factor appearing in the interaction term in $\gamma_y$;

\item
the appearance of the $\sigma^{z}$ operator (parameter) of the last spin of the triplet in the interaction terms in $\gamma_y$ and $\gamma_z$;

\item
the Pauli spin operators ($\hat{\sigma}^{x}$, $\hat{\sigma}^{y}$ and $\hat{\sigma}^{z}$) of the first spin of the triplet under consideration remain unchanged in each relative interaction term ($\gamma_x$, $\gamma_y$ and $\gamma_z$).
\end{itemize}

The final set of parametric single-spin-1/2 Hamiltonians is then
\begin{equation}\label{HNdispariSpinDiagonalizzataSM}
\begin{aligned}
\tilde{H}=\gamma_{x}\hat{\sigma}_{1}^{x}&+
\Biggl[(-1)^{N-1\over 2}\gamma_{y}\prod_{k=1}^{(N-1)/2}\sigma_{2k+1}^{z}\Biggr]\hat{\sigma}_{1}^{y} \\
&+\Biggl[\gamma_{z}\prod_{k=1}^{(N-1)/2}\sigma_{2k+1}^{z}\Biggr]\hat{\sigma}_{1}^{z},
\end{aligned}
\end{equation}
in the case of an odd number of spins, and
\begin{equation}\label{HNPariSpinDiagonalizzataSM}
\tilde{H}=\gamma_{x}\hat{\sigma}_{1}^{x}+\Biggl[(-1)^{N \over 2}\gamma_{y}\prod_{k=1}^{N/2}\sigma_{2k}^{z}\Biggr]\hat{\sigma}_{1}^{x}+\gamma_{z}\prod_{k=1}^{N/2}\sigma_{2k}^{z}.
\end{equation}
when $N$ is an even number.
For the sake of clearness, we point out that ${(N-1)/2}$ and $N/2$, appearing respectively in Eq. \eqref{HNdispariSpinDiagonalizzataSM} and \eqref{HNPariSpinDiagonalizzataSM}, are the numbers of transformations necessary to get  the final set of two-level Hamiltonians from the original $N$-spin Hamiltonian [Eq. \eqref{HNspinMessinaSM}].

Finally, we can consider local magnetic fields applied to the spins of the chain, that is a further term in the Hamiltonian in Eq. \eqref{HNspinMessinaSM} of the following type
\begin{equation}
    \sum_{k=1}^N \hbar \omega_k \hat{\sigma}_k^z.
\end{equation}
In the case of two and three spins ($k=2$ and $k=3$, respectively), the above operator, subject to the unitary transformation based on the operators \eqref{OperatoreUH2spinMessinaSM} and \eqref{OperatoreTrasfUH3spinMessinaSM}, acquires the forms
\begin{subequations}
  \begin{align}
    &\hbar(\omega_1 + \sigma_2^z \omega_2) \hat{\sigma}_1^z, \\ \nonumber \\
    \hbar(\omega_1 &+ \sigma_2^z \omega_2 + \sigma_2^z \sigma_3^z \omega_3) \hat{\sigma}_1^z,
  \end{align}
\end{subequations}
respectively.
Then, in the case of $N$ spins, the general form of the factor multiplying $\hat{\sigma}_{1}^{z}$ and depending on the $\omega_k$ parameters reads
\begin{equation}
\hbar \left[ \omega_{1}+\sum_{k=2}^{N}\prod_{k'=2}^{k} \sigma_{k'}^{z} \omega_{k} \right].
\end{equation}

\bibliography{biblio_spin.bib}

\Ignore{

}

\end{document}